\documentclass[10pt]{iopart}
\usepackage{epsfig}
\usepackage{color}
\usepackage{colortbl}
\usepackage[table]{xcolor}
\usepackage{mathrsfs}
\usepackage{amssymb}
\usepackage{graphicx}

\renewcommand{\rho}{\varrho}

\renewcommand{\e}{\mathtt{e}}
\newcommand{\im}{\mathtt{i}}
\newcommand{\p}{\partial}
\newcommand{\R}{\mathbf{R}}
\newcommand{\C}{\mathbf{C}}

\newcommand{\K}{\mathcal{K}}
\renewcommand{\d}{\mathtt{d}}
\newcommand{\x}{\tilde{x}}
\newcommand{\xx}{\bar{x}}
\newcommand{\dx}{\mathtt{d}x}

\newcommand{\dt}{\mathtt{d}t}
\newcommand{\dy}{\mathtt{d}y}
\newcommand{\ds}{\mathtt{d}s}

\newcommand{\argmin}{\mathrm{argmin}\,}
\newcommand{\argmax}{\mathrm{argmax}\,}

\newcommand{\vk}{\varkappa}

\renewcommand{\S}{{\mathscr S}}

\newcommand{\Erf}{{\mathrm{Erf}}}
\renewcommand{\P}{{\mathbb P}}

\renewcommand{\geq}{\geqslant}
\renewcommand{\leq}{\leqslant}

\newcommand{\DUE}{\mathtt{DUE}}
\newcommand{\GUE}{\mathtt{GUE}}
\newcommand{\GOE}{\mathtt{GOE}}

\newcommand{\LL}{\mathcal{L}}
\newcommand{\D}{\mathscr{D}}
\renewcommand{\theta}{\vartheta}
\newcommand{\Dom}{{\mathrm{Dom}}}

\newcommand{\BE}{\begin{equation}}
\newcommand{\EE}{\end{equation}}

\begin{document}

\title[Inner structure of vehicular ensembles and random matrix theory]{Inner structure of vehicular ensembles and random matrix theory}

\author{Milan Krb\'alek and Tom\'a\v s Hobza}

\address{Faculty of Nuclear Sciences and Physical Engineering, Czech Technical University in Prague,
Prague, Czech Republic}
\ead{milan.krbalek@fjfi.cvut.cz}

\begin{abstract}
We introduce a special class of random matrices (DUE) whose spectral statistics corresponds to statistics of microscopical quantities detected in vehicular flows. Comparing the level spacing distribution (for ordered eigenvalues in unfolded spectra of DUE matrices) with the time-clearance distribution extracted from various areas of the flux-density diagram (evaluated from original traffic data measured on Czech expressways with high occupancies) we demonstrate that the set of classical systems showing an universality associated with Random Matrix Ensembles can be extended by traffic systems.
\end{abstract}


\maketitle

\section*{Introduction}

Connection between Random Matrix Ensembles and certain transport systems is not new (see e.g.\cite{Tao} for a general overview). Indeed, in Ref.\cite{Cuernavaca,Cuernavaca2} authors drawn the attention to the bus transportation system in Cuernavaca, Mexico, where a peculiar transfer of information inside the system led, surprisingly, to the universal configuration of gaps among buses. The rigorous study \cite{Deift} confirmed that the detected link between Cuernavaca buses and Gaussian unitary ensemble ($\GUE$) is not accidental. Authors of the research formulated a relevant microscopic model-scheme and analytically proved that the time headway distribution of buses conforms to the level spacing distribution for $\GUE$. This surprising knowledge attracted an attention of scientists (e.g. \cite{Tao}) since systems connected to Random Matrix Theory are universal (in a certain sense).

Until now, natural endeavor to interconnect Random Matrix Theory (RMT) with vehicular systems has not led to a success. However, a partial progress has been achieved in \cite{Helbing_and_Krbalek,Red_cars,Traffic_NV,My_Multiheadways,Red-intervals} where it is proved that microscopical arrangement of vehicles can be predicted by means of a certain one-dimensional gas inspired by the Dyson's gases that are well-known in RMT. Moreover, another attempts to describe (analytically) a microstructure of vehicular ensembles with help of statistical instruments (e.g. \cite{May,Jin_and_Zhang,Li_and_Wang}) led us to believe that our effort may be successful. Therefore, the main goal of this work is to find (and analyze) a new class of random matrices whose spectral properties correspond to a micro-structure of real-road traffic samples.

This paper is organized as follows. In the first section we introduce and analyze the class of the so-called damped matrices ($\DUE$). We are focused on the level density, the procedure of the unfolding, and the level spacing distribution. Connection to the theory of one-dimensional traffic gases is discussed in the second section. Section Three brings in a comparison between  the level spacing distribution of $\DUE$ matrices and the time-clearance distributions extracted from expressway data samples. This is followed by corresponding discussion and conclusions in the final section. Subsidiary derivations are included in the mathematical Appendix.

\section{DUE -- Damped unitary ensembles}

Being inspired by the work \cite{Bogomolny-CM-matrices} (studying Calogero-Moser models with various potentials) we introduce the $g-$parameterized class $\DUE_g(N)$ whose $N\times N$ matrices $H_g=(h_{kj}(g))_{k,j=1}^N$ fulfil the following axioms:

\begin{enumerate}
\item the parameter $g\geq 0$ is fixed;
\item elements $h_{kk}(g)$ are chosen as independent Gaussian variables with zero mean and unit variance, i.e. $h_{11}(g),h_{22}(g),\ldots,h_{NN}(g) \backsim \mathrm{N}(0,1)$ are i.i.d.;
\item if $k\neq j$ then $$h_{kj}(g)=\frac{2\pi\im g}{N \sinh[(2\pi(k-j)/N]},$$ i.e. the off-diagonal elements are deterministic.
\end{enumerate}

Owing to the definition the off-diagonal elements are purely imaginary complex numbers whose absolute value become smaller with distance from the diagonal. Therefore we refer such a class to as the Damped Unitary Ensembles. In the context of the article \cite{Bogomolny-CM-matrices} the matrices in $\DUE_g(N)$ represent simplified versions of the $N \times N$ Lax matrices derived for hyperbolic Calogero-Moser model (see \cite{Bogomolny-CM-matrices} for details).

\subsection{Level density for $\DUE$ matrices}

\begin{figure}[htbp]
\begin{center}
\hspace*{10mm}\includegraphics[width=9cm, angle=0]{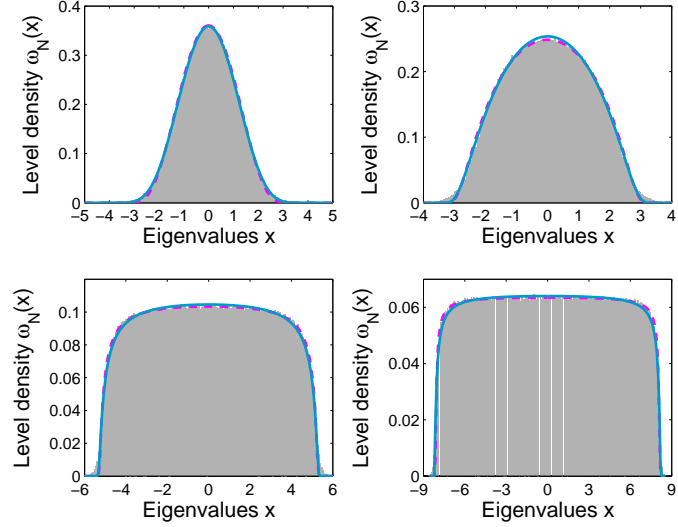}
\end{center}
\caption{Graphs of the level density $\omega_N(x).$ Histograms have been determined for five thousand realizations of $\DUE(256)$ matrices with parameters $g=0.2$ (a northwest sub-figure), $g=0.5$ (a northeast sub-figure), $g=1.5$ (a southwest sub-figure), and $g=2.5$ (a southeast sub-figure).    Solid curve (blue)/dashed curve (magenta) represent the approximation (\ref{pilgrim}) for values $\varepsilon,$ $\theta$ estimated with help of CSC/KE methods, respectively. \label{fig:DUE-LD-ctyrobrazek}}
\end{figure}

Let $\sigma(H_g)$ be the spectrum of $H_g,$ i.e. $\sigma(H_g)=\{x_{(k)}\in\C:k=1,2,\ldots,N\}$ is a set of ordered eigenvalues $x_{(1)} \leq x_{(2)} \leq \ldots \leq x_{(N)}.$  Since $H_g=H^\sharp_g,$ where $H^\sharp_g$ is Hermitian conjugated matrix, one finds that $\sigma(H_g)\subset\R$ and therefore the above-referred ordering is correctly defined. Now we can introduce the probability density $P_k(x_{(k)})$ for $k$th eigenvalue $x_{(k)}$ and the level density
\BE \omega_N(x)=\frac{1}{N}\sum_{k=1}^N P_k(x). \EE
Unfortunately, the famous Semi-Circle Law (derived in \cite{Mehta} for the level density of classical random matrix ensembles) is not applicable for $\DUE$ matrices. Instead of a circular shape the level density of damped matrices conforms to a probability density taken from two-parametric family
\BE   q_{\theta,\varepsilon}(x)=\frac{\zeta(\theta)}{\varepsilon}\left\{\begin{array}{ccc} \exp[-\frac{\theta^2\varepsilon^2}{\varepsilon^2-x^2}]& \ldots& |x| < \varepsilon,\\ 0 & \ldots& |x| \geq \varepsilon, \end{array}\right. \label{pilgrim} \EE
where $\zeta^{-1}(\theta)=\int_{-1}^1 \exp[-\frac{\theta^2}{1-x^2}]\,\dx$ ensures the proper normalization. Indeed,  numerical tests show (as is illustrated in figure \ref{fig:DUE-LD-ctyrobrazek}) that the level density for large $N$ is very accurately approximated by the function $q_{\theta,\varepsilon}(x)$ for estimated values $\hat{\varepsilon}=\hat{\varepsilon}(g,N),\hat{\theta}=\hat{\theta}(g,N)$  of the so-called critical eigenvalue $\varepsilon$ and curvature $\theta.$ Both the presented estimation-procedures are based on principles of MDE (minimum distance estimation). The first of them (CSE -- chi-square estimator) minimizes the statistical distance
\BE \chi(\theta,\varepsilon)=\sqrt{\int_{-\infty}^{+\infty} |h(x)-q_{\theta,\varepsilon}(x)|^2\,\dx},\EE
where $h(x)$ is an empirical histogram. For potential reconstructive purposes we are specifying that the set $S$ consisting of $1\, 280\,000$ eigenvalues (taken from $5\,000$ realizations of $256\times256$  matrices) generates the domain $\Dom(h)=[\min(S),\max(S)]$ of the empirical histogram. In our numerical tests, this domain $\Dom(h)$ has been divided into $200$ equidistant sub-intervals.

\begin{figure}[htbp]
\begin{center}
\hspace*{10mm}\includegraphics[width=9cm, angle=0]{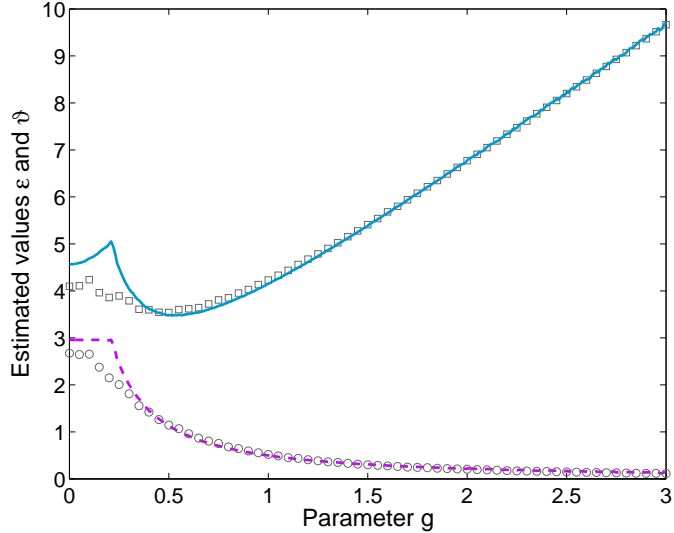}
\end{center}
\caption{Estimates of the critical eigenvalue $\varepsilon$ and curvature $\theta$ in (\ref{pilgrim}). A curve and dashed line correspond to $\hat{\varepsilon}$ and $\hat{\theta}$  obtained by CSE method. Squares and circles display the estimates $\hat{\varepsilon}$ and $\hat{\theta}$ calculated by means of KE.      \label{fig:DUE-LD-parameters}}
\end{figure}

The second statistical test is based on the standard Kolmogorov estimator (KE) minimizing the supremum of the absolute difference between the estimated distribution function and the empirical distribution function
\BE H_N(x) = \frac{1}{N}  \sum_{k=1}^{N} I(x_{(k)} \leq x).\EE
Here, $I(x)$ stands for the indicator function. The optimal values of estimated parameters are compared graphically in figure \ref{fig:DUE-LD-parameters}. Marked discrepancies near the origin can be attributed to the following facts. If $g=0$ then one can trivially express the level density as $\e^{-\frac{x^2}{2}}/\sqrt{2\pi}$ and our estimation procedure tries, in fact, to approximate the normal distribution by a function $q_{\theta,\varepsilon}(x).$ Moreover,
$$\forall \theta>0: \quad \lim_{\varepsilon\rightarrow0_+} q_{\theta,\varepsilon}(x)=\lim_{\varkappa\rightarrow0_+} \frac{1}{\sqrt{2\pi}\varkappa} \e^{-\frac{x^2}{2\varkappa^2}}=\delta(x),$$
where $\delta(x)\in\D'$ is the Dirac function (see Appendix \ref{subsec:APP-01}). Since both these distributions tend (for small values of parameters) to the same generalized limit it is difficult to distinguish them numerically. It means that estimations used (applied for small values of $g$) are highly sensitive on a type of estimators, quality of analyzed data and so on. On the other hand, nuances in parameters estimated by different methods cause only imperceptible deviations in curves of the level density (see solid and dashed lines in the northwest sub-figure in figure \ref{fig:DUE-LD-ctyrobrazek})).

\subsection{Procedure of the unfolding}\label{subsec:UNFO}

Before statistical analysis of random-matrix spectra the standard methodology of Random Matrix Theory requires a transition to the so-called \emph{unfolded spectrum.} The main benefit of that transition is an unified view into a structure of heterogeneous systems. To be specific, the procedure of the unfolding converts eigenvalue-spectrum $\sigma(H_g)$ with an arbitrary level density $\omega_N(x)$ (depending on specific properties of a system) to the unfolded spectrum $\tilde{\sigma}(H_g)=\{\x_{(k)}\in\C:k=1,2,\ldots,N\}$ in which unfolded eigenvalues are uniformly distributed in $(0,N).$ Thus, the level density of the unfolded spectrum is associated with the uniform distribution $U(0,N).$ Mathematically, the unfolding is a procedure consisting of two operations: an individual mapping $x_{(k)} \mapsto \xx_{(k)}$ defined by
\BE \xx_{(k)}:=\int_{-\infty}^{x_{(k)}} \omega_N(y)\,\dy,\label{UNFO}\EE
and a global scaling
\BE \x_{(k)}:=\frac{\xx_{(k)}(N-1)}{\sum_{k=1}^{N-1} (\xx_{(k+1)}-\xx_{(k)})}.\EE According to the theorem on probability integral transformation (see \cite{Oxford}) it holds $\xx \backsim \mathrm{U}(0,1),$ which results (for large $N$) in the fact $\x \backsim \mathrm{U}(0,N).$ Moreover, the average gap between two succeeding unfolded eigenvalues is equal to one.

\emph{Specifications:} All the previous considerations require a large value of $N.$ Thus, in our numerical realizations we consistently use $N=256.$

\subsection{Level spacing distribution}\label{sec:DUE-LS}

In this subsection we aim to deduce an analytical formula predicting the probability density for gaps $s_k:=\x_{(k+1)}-\x_{(k)}$ between the ordered eigenvalues $\x_{(1)} \leq \x_{(2)} \leq \ldots \leq \x_{(N)}$ in the unfolded spectrum $\tilde{\sigma}(H_g).$ Here, such a function is referred as the \emph{level spacing distribution (LS-distribution)} and denoted by $\wp(s).$

The most trivial variant of ensembles investigated is the class $\DUE_0(N)$ of totally damped matrices. The matrices $H\in\DUE_0(N)$ are diagonal and therefore the corresponding level density reads $\omega_N(x)=\frac{1}{\sqrt{2\pi}} \e^{-x^2/2}$ for all $N.$ Thus, the first stage of the unfolding-procedure
\BE \xx:=\frac{1}{\sqrt{2\pi}} \int_{-\infty}^x \e^{-y^2/2}\,\dy= \frac{1}{2}\left(1+\Erf[\frac{x}{\sqrt{2}}]\right) \EE
produces i.i.d.-numbers from the uniform distribution $U(0,1).$ Hence, one can easily verify that level spacing distribution in $\DUE_0(N)$ can be expressed as
\BE \wp(s)=\Theta(s)\e^{-s}. \label{DIAGONAL-LS} \EE

In the article \cite{Bogomolny-CM-matrices} authors derived that the joint probability density for eigenvalues in the hyperbolic Calogero-Moser model corresponds to the law
\BE P(x_1,x_2,\ldots,x_N) \propto \exp\bigl[-\Omega\sum_{k=1}^N x_k^2\bigr]\prod_{k=1}^N \K_0\left(\Psi\prod_{j\neq k} \left|1+\frac{4\pi\im g}{N(x_j-x_k)}\right|\right), \label{CMH-joint} \EE
where $\Omega,\Psi$ are constants and $\K_m(x)$ stands for the Macdonald's function of the order $m$ -- solution of the modified Bessel's differential equation of the second kind $x^2y''+xy'-(x^2+m^2)y=0.$ Having regard to the fact that $\DUE_g(N)$ matrices can be understood as numerical simplifications of Lax matrices in the hyperbolic Calogero-Moser model (as is discussed in \cite{Bogomolny-CM-matrices}), the function (\ref{CMH-joint}) can be considered as a suitable approximation for the joint probability density in the ensemble $\DUE_g(N).$ Although the expression (\ref{CMH-joint}) is difficult to handle, it allows to obtain a Wigner-type surmise for the level spacing distribution in $\DUE_g(N).$ Considering $N=2$ (similarly to the original Wigner idea) we have
\BE P(x_1,x_2) \propto \e^{-\Omega(x_1^2+x_2^2)} \K_0^2 \left(\Psi\sqrt{1+\frac{4\pi^2g^2}{(x_2-x_1)^2}}\right). \EE
From the general relation $\ell(r)=\Theta(r)\left(\int_\R P(r+x,x)\,\dx+\int_\R P(x,r+x)\,\dx\right)$ we can obtain the probability density %
\BE \ell(r) \propto \Theta(r) \K_0^2 \left(\Psi\sqrt{1+\frac{4\pi^2g^2}{r^2}}\right) \e^{-\Omega r^2/2} \EE
for a distance between non-unfolded eigenvalues in $\DUE_g(2).$ Application of the approximation $\K_0(z) \approx  \sqrt{\frac{\pi}{2z}}\e^{-z} + \mathcal{O}(1/z)$ is resulting in the formula
\BE \ell(r) \propto r \e^{-\frac{4\pi \Psi g}{r}}, \label{level-repulsion}  \EE
describing a behavior of the distribution studied at small distances $r.$ An immediate consequence of this result is evidence of  repulsions between eigenvalues that are rapidly stronger than those detected in the standard matrix ensembles (i.e. $\GOE,$ $\GUE$).

Combining the two asymptotic predictions (\ref{DIAGONAL-LS}) and (\ref{level-repulsion}) we suggest (analogously to \cite{Bogomolny-CM-matrices}) the level spacing distribution in the form
\BE \wp(s)=A s^\alpha \e^{-\beta/s-Ds}, \label{LS-distribution}\EE
where (according to Appendix \ref{subsec:APP-02})
\BE D=\beta+\alpha+\frac{3-\e^{-\sqrt{\beta}}}{2}, \label{D-scaling-constant}\EE
\BE A^{-1}=2 \left(\frac{\beta}{D}\right)^{\frac{\alpha+1}{2}}\K_{\alpha+1}(2\sqrt{\beta D}). \label{A-scaling-constant} \EE

\begin{figure}[htbp]
\begin{center}
\hspace*{10mm}\includegraphics[width=9cm, angle=0]{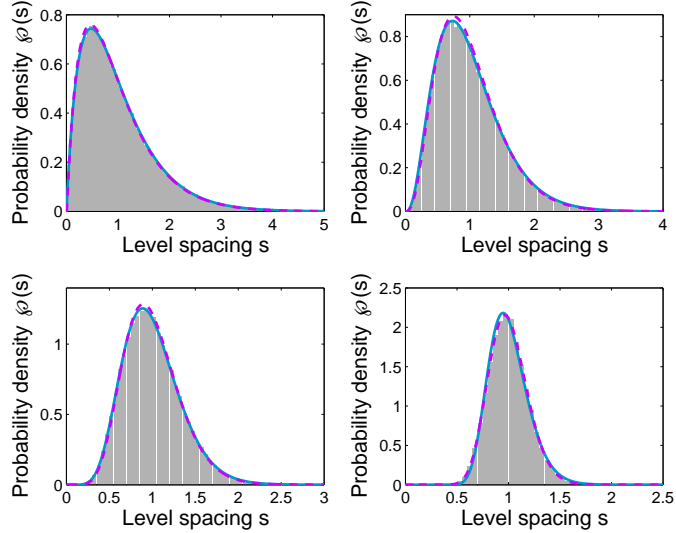}
\end{center}
\caption{Level spacing distribution. The gray histograms correspond to empirical LS-distributions of $\DUE_g(256)$ obtained for one thousand matrix-realizations and for $g=0.2$ (a northwest sub-figure), $g=0.5$ (a northeast sub-figure), $g=1.0$ (a southwest sub-figure), and $g=2.0$ (a southeast sub-figure). The continuous/dashed curves display the prognosis (\ref{LS-distribution}), in which the parameters have been estimated by means of MDE/MLE estimators, respectively. For all variants the value $m=50$ has been used.  \label{fig:DUE-LS-ctyrobrazek}}
\end{figure}

\begin{figure}[htbp]
\begin{center}
\hspace*{10mm}\includegraphics[width=9cm, angle=0]{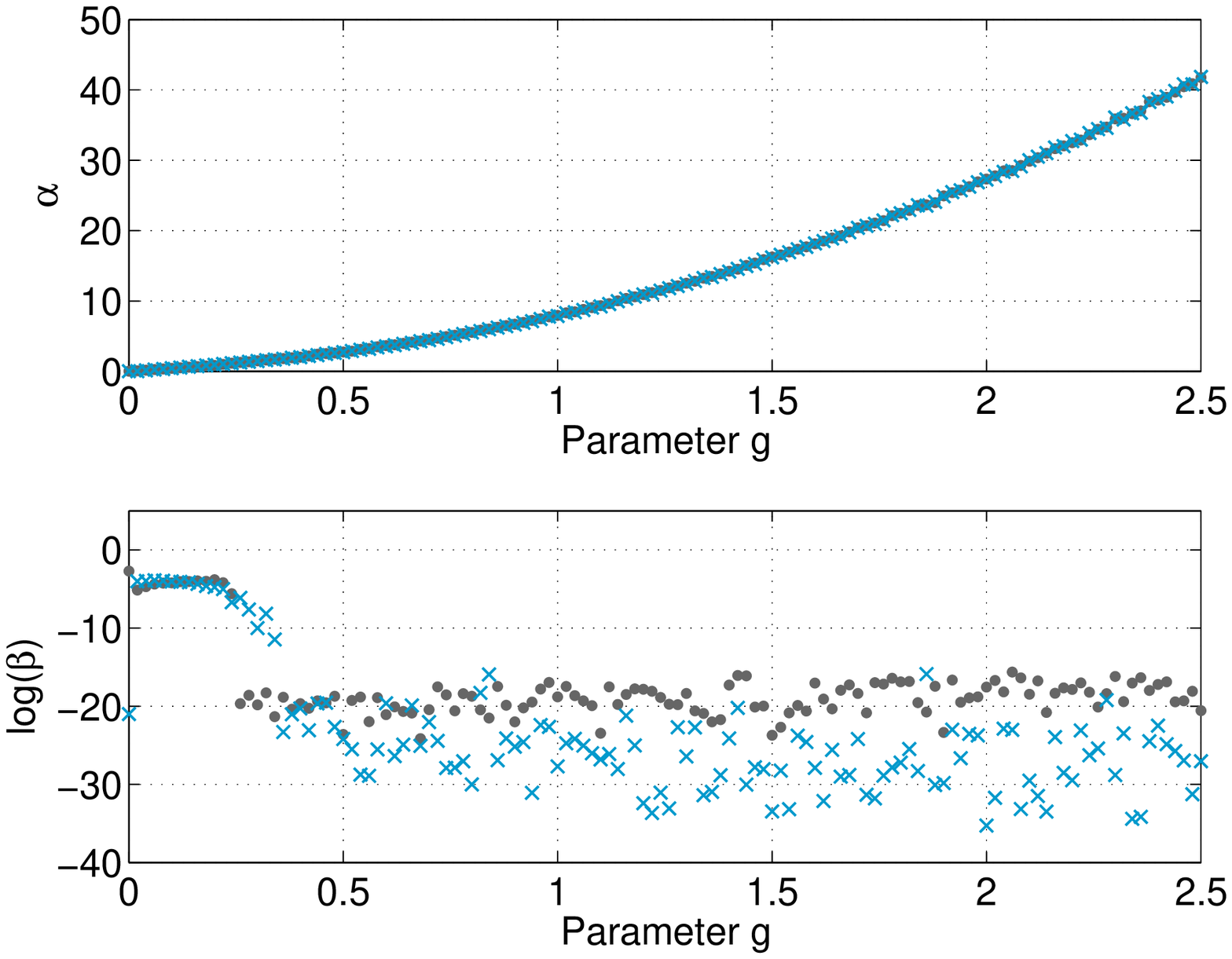}
\hspace*{10mm}\includegraphics[width=9cm, angle=0]{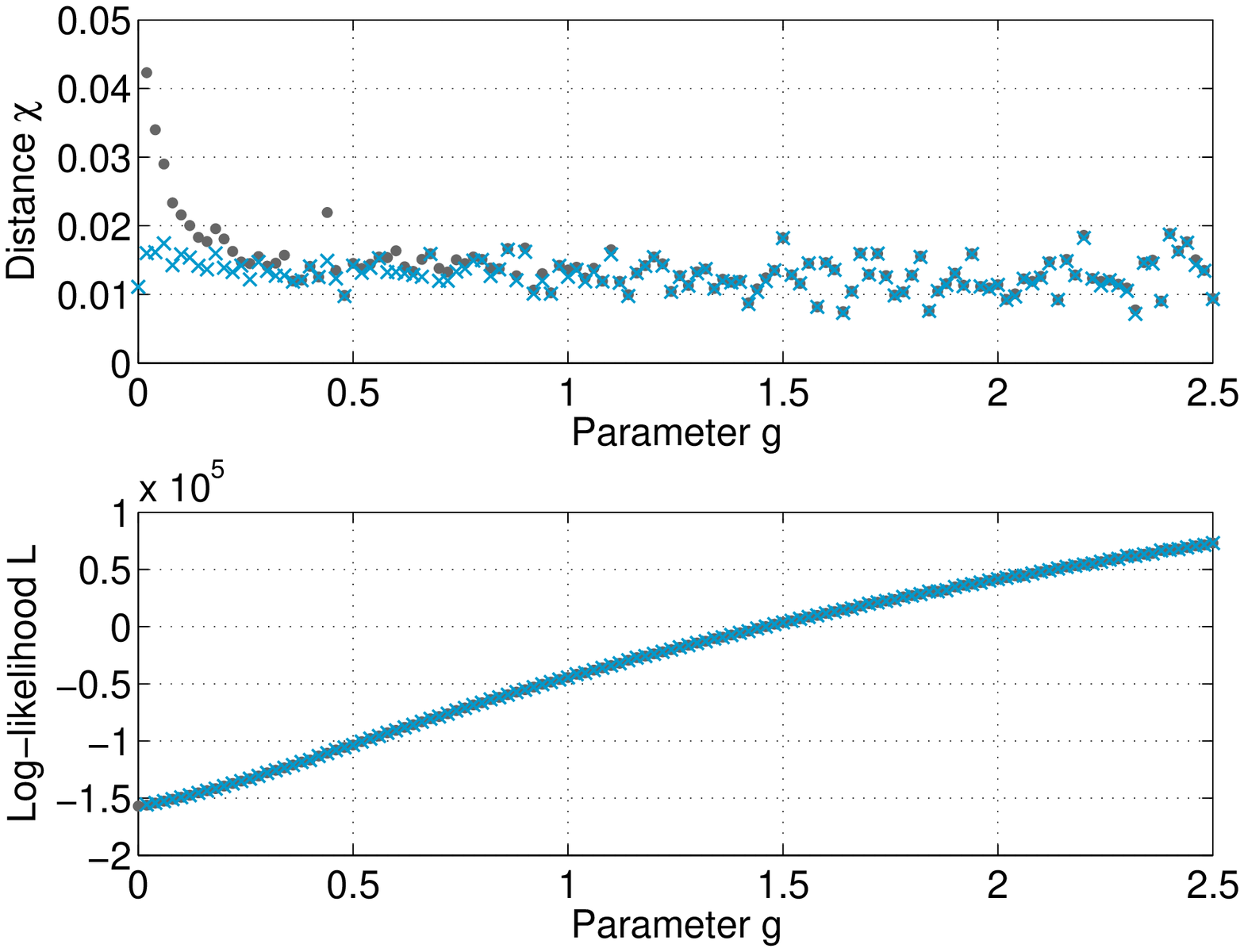}
\end{center}
\caption{Estimated values of parameters in the functional premise (\ref{LS-distribution}). We visualize the parameters $\alpha,\beta$ estimated by the maximum likelihood estimation (bullets) and minimum distance estimation (crosses), respectively. Values of the statistical distance $\chi(\alpha,\beta)$ between the probability density (\ref{LS-distribution}) specified for estimated parameters and empirical histograms are plotted in the third sub-figure. Values of the log-likelihood $\LL(\alpha,\beta)$ corresponding to the probability density (\ref{LS-distribution}) specified for estimated parameters are plotted in bottom sub-figure. \label{fig:DUE-LS-parameters}}
\end{figure}

On contrary to the pioneering work \cite{Bogomolny-CM-matrices} our paper intends to investigate the level spacing distribution of $\DUE$-matrices in more details. As the curse of the level density near the values $x=\pm\varepsilon$ is extremely steep (see, for example, the southeast sub-figure \ref{fig:DUE-LD-ctyrobrazek}), the numerical representation of the unfolding (using an approximate formula (\ref{pilgrim}) for the level density) is unstable in those regions. For this reason we analyze a truncated spectrum $\tilde{\sigma}_{m}(H_g)=\{\x_{(k)}:~k=m+1,2,\ldots,N-m\}$ whose elements are re-scaled so that the mean spacing is equal to one. Then the set $\{s_{m+1},s_{m+2},\ldots,s_{N-m-1}\}$ of empirical spacings  is subjected to standard procedures estimating optimal values $\alpha,\beta$ in the functional premise (\ref{LS-distribution}). Here we present two standard estimators: MLE -- maximum likelihood estimation and MDE -- minimum distance estimation. For these purpose we introduce the log-likelihood function
\BE \LL(\alpha,\beta)=(N-2m-1)\ln(A)+\alpha\sum_{k=m+1}^{N-m-1} \ln(s_k) - \beta \sum_{k=m+1}^{N-m-1} \frac{1}{s_k} - D \sum_{k=m+1}^{N-m-1} s_k  \EE
and the weighted statistical distance
\BE \chi(\alpha,\beta):=d[h,\wp]=\sqrt{\int_0^{+\infty} \frac{\pi s}{2}\e^{-\frac{\pi s^2}{4}} \bigl|h(s)-\wp(s;(\alpha,\beta))\bigr|^2\,\ds},\label{vahova-vzdalenost}\EE
where $h(s)$ is an empirical histogram for spacings. We remark that the weight factor $\varpi(s):=\frac{\pi s}{2}\e^{-\frac{\pi s^2}{4}}$ has been chosen to eliminate the influence of long spacings and increase the influence of spacings that are close to the mean value. In addition, $\argmax_{s \geq 0} \varpi(s)= 1$ and $d[0,1]=1,$ which can be understood (if $d$ is now understood as a general functional metric) as a benchmark. As graphically illustrated in figure \ref{fig:DUE-LS-ctyrobrazek} LS-distributions of $\DUE-$matrices is in full compliance with the analytical prediction (\ref{LS-distribution}). The optimal values of estimated parameters, i.e. $(\hat{\alpha},\hat{\beta})_{\mathrm{MLE}}=\argmax \LL(\alpha,\beta),$ $(\hat{\alpha},\hat{\beta})_{\mathrm{MDE}}=\argmin \chi(\alpha,\beta)$ are then compared graphically in figure \ref{fig:DUE-LS-parameters}.

\section{Local thermodynamical traffic model}\label{sec:traffic-model}

As one of the main goals of this article is to find a connection between the above-referred matrix ensembles and freeway traffic samples, it is beneficial to show that the functional premise (\ref{LS-distribution}) can be understood as a steady-state distribution of particle headways in the thermal-like socio-physical traffic model with
short-ranged repulsions between particles (originally introduced in \cite{Krbalek_gas}). Indeed, if considering circular ensemble of  particles whose mutual distances (between subsequent particles) are denoted by $s_1,s_2,\ldots,s_n$ and if understanding such an ensemble as thermodynamical (with a thermal parameter $\nu\geq 0$) we are able to derive (following the approaches presented in \cite{Krbalek_gas}) that the headway distribution in the ensemble with a repulsive potential energy
\BE U_\mu(s_1,s_2,\ldots,s_n)=\sum_{k=1}^n \left(\frac{1}{s_k}-\mu\ln(s_k)\right), \quad (\mu \geq 0) \label{snow}\EE
takes the form $\wp(s)=As^{\mu\nu}\e^{-\frac{\nu}{s}}\e^{-Ds}.$ Such a probability density fully corresponds to the generalized inverse Gaussian distribution distribution (\ref{LS-distribution}) and moreover, it fulfills all the substantial criteria for acceptability of analytical clearance distributions presented in the articles \cite{My_Multiheadways,Red-intervals}. Indeed, for all values of $\mu\geq 0$ and all $\nu>0$ the function $\wp(s)$ generates the origin plateau, which means that it obeys the condition $q\in\R^+ \Rightarrow \lim_{s\rightarrow 0_+} s^{-q}\wp(s)=0.$ Furthermore, the function $\wp(s)$ is belonging to the class of balanced distributions (see \cite{Red-intervals}), which (together with the mathematical criteria \textbf{(T1--T5)} discussed in \cite{Red-intervals}) completes all theoretical requirements for vehicular headway distributions.

Since the presented thermodynamical traffic model represents a mathematically well-defined (and analytically soluble) alternative to the more realistic traffic model (see the discussion on the pages 6,7 in \cite{My_Multiheadways}) we have now a theoretical permission to use the above-referred statistics for a mathematical description of headway distributions in realistic vehicular streams.

\section{Headway statistics in various regions of flux-density map}

In this section we try to answer the fundamental question of this work: "Is there any link between fluid vehicular locations and eigenvalues in random matrices (similar to the link between Mexico buses and $\GUE-$matrices presented in \cite{Cuernavaca,Cuernavaca2,Deift,Tao})?"

\subsection{On the origin of traffic data used}

Vehicle-by-vehicle data analyzed in the paper have been recorded at the Expressway R1 (also called the Prague Ring) in Prague, the Czech Republic. For our measurements (conducted in cooperation with The Road and Motorway Directorate of the Czech Republic) we have chosen expressway-segments where traffic is strongly saturated and where therefore one can regularly detect synchronized flows. Having regard to intentions of this article we eliminate all free-flow states (where car-car interactions are negligible and therefore traffic behaves almost like poissonian system). For distinguishing congested-traffic states from free-flow states (see \cite{Martin,Review-Kerner}) we have applied the clustering algorithm K-means (see \cite{K_means}). Here the chosen alternative of the method is based on minimization with respect to velocity of vehicles.

\subsection{The traffic alternative for the spectral unfolding}

Data record (adapted into a form suitable for further processing) has been divided into sub-samples $T_j=\{t_{j1},t_{j2},\ldots,t_{jm}\}$ containing $m$ consecutive time clearances (netto time intervals between succeeding cars passing a traffic detector located at the chosen lane), where $m$ is the sampling size. For each sub-sample there were calculated the local flux $J_j$ and local density $\varrho_j.$ The method used for those calculations is identical to the standard method described mathematically in \cite{My_Multiheadways} (page 4). Again, we emphasize that all sub-samples belong to a regime of congested streams when average speed of cars is dramatically reduced compared to the maximum allowable speed.

In order to avoid an undesirable mixing between different traffic constellations we apply (similarly to the procedure presented for random matrices in the subsection \ref{subsec:UNFO}) the so-called \emph{traffic unfolding.} Here, such a procedure is composed of two components: a \emph{re-scaling} and \emph{segmentation} by a flux-density window. To be specific, for a certain window-size $(\Delta_J,\Delta_\varrho)$ we introduce a flux-density window $W(\varrho,J):=[\varrho,\varrho+\Delta_\varrho]\times[J,J+\Delta_J]$ that represents a small rectangular sub-region inside the flux-density map. Then the re-scaling is understood as a transformation of the set $T_j$ into the set $\{\tau_{j1},\tau_{j2},\ldots,\tau_{jm}\},$ where $\tau_{jk}=t_{jk}/\sum_{k=1}^m t_{jk}.$ Furthermore, the segmentation
\BE I(\varrho,J)=\{j:\,J_j\in [J,J+\Delta_J]\,\wedge\, \varrho_j\in [\varrho,\varrho+\Delta_\varrho]\}\EE
is understood as a procedure selecting sub-samples associated with a chosen flux-density window $W(\varrho,J).$ It means that the final sample of unfolded clearances looks like
\BE \mathcal{T}(\varrho,J)=\{\tau_{jk}:\,j\in I(\varrho,J)\,\wedge\,k=1,2,\ldots,m\}. \label{unf-sets}\EE

For completeness, we add that in this study there is consistently considered $m = 50,$ $\Delta_J=400\,veh/h,$ and $\Delta_\varrho=5\, veh/km.$ The total number of analyzed clearances is circa 45 millions (7 record-files gauged on 7 different detectors between 2013-09-16 and 2014-05-29). Among 45 millions of measured clearances 14 millions have been classified (with help of the K-means clustering algorithm) as congested (or transitional).

\begin{figure}[htbp]
\begin{center}
\hspace*{10mm}\includegraphics[width=9cm, angle=0]{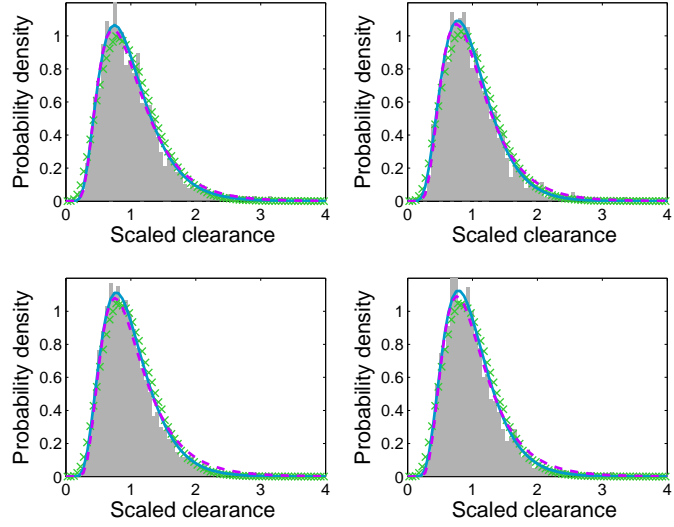}
\end{center}
\caption{Time clearance distributions in vehicular flows (analyzed for fast lanes of expressways). Histograms stand for empirical distributions analyzed inside the following flux-density windows: a northwest sub-figure: $W(48,2300),$ northeast sub-figure: $W(54,1900),$ southwest sub-figure: $W(58,1300),$ and southeast sub-figure: $W(62,1300).$ The continuous/dashed curves visualize functions (\ref{TC-distribution}) specified for parameters estimated by MDE/MLE, respectively. The crosses correspond to a nearest distribution of eigenvalue-gaps, i.e. to a level spacing distribution whose statistical distance to traffic histogram is minimum. \label{fig:LS-traffic}}
\end{figure}

\subsection{Changes in headway statistics induced by varying fluxes and densities}

In this subsection we aim to analyze a clearance distribution extracted from unfolded data-sets (\ref{unf-sets}) associated to a fixed flux-density window $W(\varrho,J).$ Withal we try to test a hypothesis that clearance distributions detected in regions of congested traffic belong to the two-parametric family of distributions
\BE \wp(\tau)=A \tau^\alpha \e^{-\beta/\tau-D\tau}. \label{TC-distribution}\EE
If proving that hypothesis we find a direct link between inner structures of vehicular ensembles and spectra of damped random matrices.

\begin{figure}[htbp]
\begin{center}
\hspace*{10mm}\includegraphics[width=9cm, angle=0]{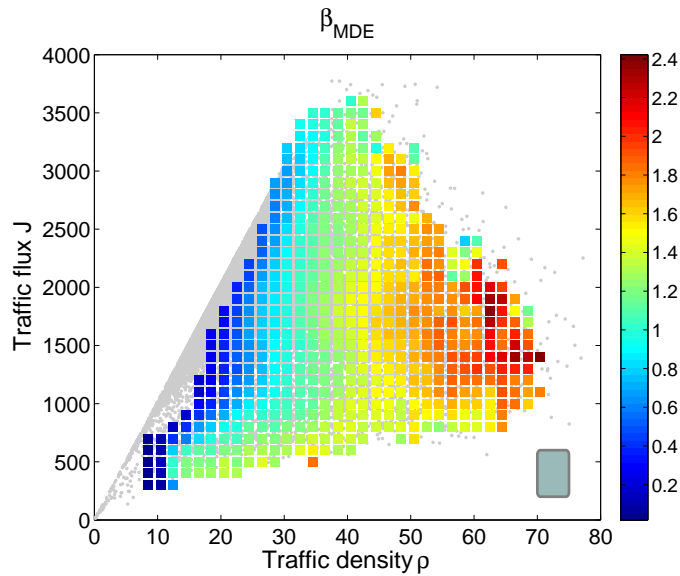}
\end{center}
\caption{Color representation of values $\beta$ estimated by minimum distance estimation (\ref{MDE}). We plot values $\hat{\beta}_{\mathrm{MDE}}$ in various flux-density windows $W(\varrho,J).$ Squares symbolize locations of window-centroids  $(\varrho+\Delta_\varrho/2,J+\Delta_J/2).$ The rectangle at the bottom right schematically visualizes the window-size used. Gray dots correspond to locations $(\rho_k,J_k)$ of individual data sub-samples before a separation procedure. Unstained areas indicate either a lack of data or belonging to a free-traffic regime. The plotted values are valid for a fast lane. \label{fig:estimated-beta-MDE}}
\end{figure}

The testing mechanism is as follows. Varying a flux-density window we estimate a optimal parameters $\alpha,\beta \geq 0$ for which the statistical distance
\BE \chi(\alpha,\beta):=d[\wp,q]=\sqrt{\int_0^{+\infty} \frac{\pi \tau}{2}\e^{-\frac{\pi \tau^2}{4}} \bigl|\wp(\tau;(\alpha,\beta))-q(\tau)\bigr|^2\,\d\tau}\label{vahova-vzdalenost-2}\EE
between a function (\ref{TC-distribution}) and clearance histogram $q(\tau)$ is minimum. Practically, after eliminating outliers $\tau_{jk}>6$ and after a repeated re-scaling (so that the average clearance is equal to one again) we solve a minimization problem
\BE (\hat{\alpha},\hat{\beta})_{\mathrm{MDE}}=\argmin_{\alpha,\beta} \int_0^6 \tau\e^{-\frac{\pi \tau^2}{4}} \bigl|\wp(\tau;(\alpha,\beta))-q(\tau)\bigr|^2\,\d\tau, \label{MDE}\EE
where the histogram $q(\tau)$ is enumerated for the data set $\tilde{\mathcal{T}}(\varrho,J)=\{\tilde{\tau}_i:\,i=1,2,\ldots,M\},$ for which $\langle  \tilde{\mathcal{T}}(\varrho,J) \rangle =1.$ Analogously, the maximum likelihood estimations were carried out. For these aims we have defined an estimator
\BE \LL(\alpha,\beta)=M\ln(A)+\alpha\sum_{i=1}^M \ln(\tilde{\tau}_i) - \beta \sum_{i=1}^M \frac{1}{\tilde{\tau}_i} - D \sum_{i=1}^M \tilde{\tau}_i.  \EE
By a maximization
\BE (\hat{\alpha},\hat{\beta})_{\mathrm{MLE}}=\argmax_{\alpha,\beta} \LL(\alpha,\beta) \EE
we obtain alternative values of estimated parameters.

\begin{figure}[htbp]
\begin{center}
\hspace*{10mm}\includegraphics[width=9cm, angle=0]{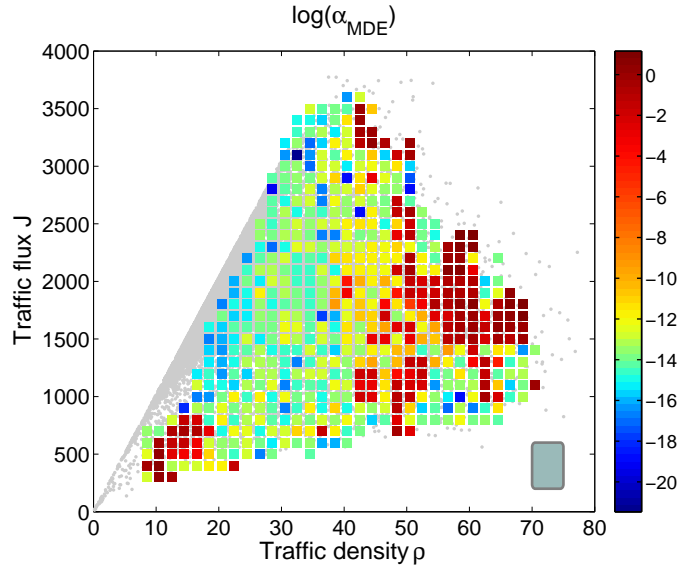}
\end{center}
\caption{Color representation of values $\alpha$ estimated by minimum distance estimation (\ref{MDE}). We plot values $\log(\hat{\alpha}_{\mathrm{MDE}})$ in various flux-density windows $W(\varrho,J).$ Squares symbolize locations of window-centroids  $(\varrho+\Delta_\varrho/2,J+\Delta_J/2).$ The plotted values are valid for a fast lane. \label{fig:estimated-alfa-MDE}}
\end{figure}

\begin{figure}[htbp]
\begin{center}
\hspace*{10mm}\includegraphics[width=9cm, angle=0]{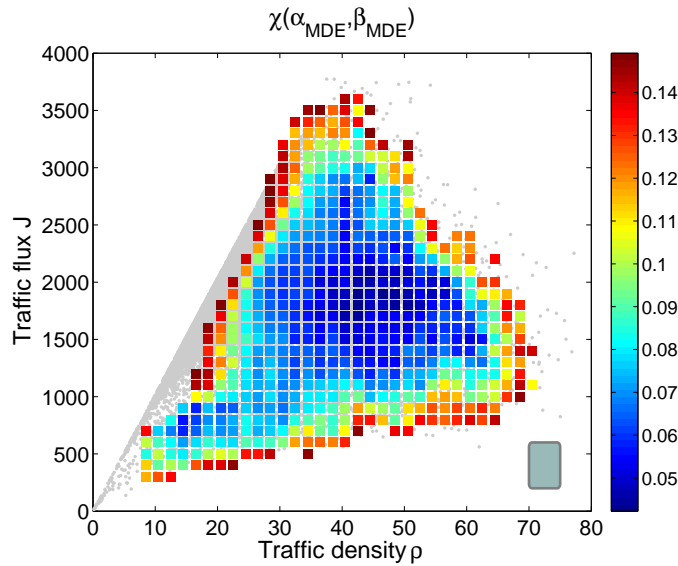}
\end{center}
\caption{Color representation of a minimum statistical distance $\chi(\alpha,\beta)$ obtained by the MDE-method. We plot values $\chi(\hat{\alpha}_{\mathrm{MDE}},\hat{\beta}_{\mathrm{MDE}})$ in various flux-density windows $W(\varrho,J).$ Squares symbolize locations of window-centroids  $(\varrho+\Delta_\varrho/2,J+\Delta_J/2).$ The plotted values are valid for a fast lane. \label{fig:deviations-TCL-MDE}}
\end{figure}

In several figures we now depict selected results of our investigations. Method MDE provides estimations visualized in figures \ref{fig:estimated-beta-MDE} and \ref{fig:estimated-alfa-MDE}. Here one can recognize a behavior so typical for agent systems with intensive interactions among agents. Indeed, with increasing density a level of vehicular coordinations is growing up. This effect is clearly visible in figure \ref{fig:estimated-beta-MDE} where estimated parameters $\hat{\beta}$ are continuously changing from negligible values (for regimes with small densities) to extremely high values detected for a strongly saturated traffic. Such a behavior fully corresponds with results published in the previous researches \cite{My_Multiheadways,Red_cars,Red-intervals,Krbalek_gas}. Since the parameter $\beta$ corresponds (in the thermodynamic model interpretation) to the inverse temperature $\nu$ discussed in the section \ref{sec:traffic-model} we can now provide a deeper insight into changes of a vehicular microstructure. Contrary to the previous investigations, in this paper we present (for the first time, as we believe) a more-detailed study of time-clearance distribution that is performed not in density windows only (as for example in \cite{Krbalek_gas,My_Multiheadways}) but in flux-density windows. The presented analysis shows that the principal quantity determining a shape of headway distribution is the traffic density. Deviations caused by changes of a flux $J$ (if density $\varrho$ is fixed) are almost imperceptible. Moreover, estimated values of $\alpha$ are low which confirms the results published formerly. If analyzing the minimum distance $\chi(\hat{\alpha},\hat{\beta})$ we can see (figure \ref{fig:deviations-TCL-MDE}) that the tested hypothesis is confirmed predominantly in the region of pure congested flows. A red triangle (boundary of a color area in figure \ref{fig:deviations-TCL-MDE}) characterizing windows $W(\varrho,J)$ with the highest deviations can be explained as follows. The east and south sides of the triangle indicate regions where the number of data is not sufficient (and histograms are not  therefore smooth enough), whereas the west side represents windows where congested traffic is mixed with free-flow traffic.

\begin{figure}[htbp]
\begin{center}
\hspace*{10mm}\includegraphics[width=9cm, angle=0]{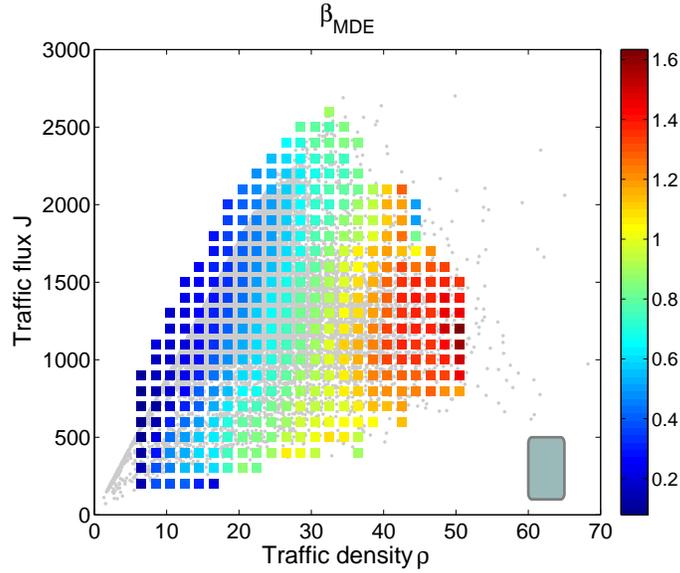}
\end{center}
\caption{Color representation of values $\beta$ estimated by minimum distance estimation (\ref{MDE}). We plot values $\hat{\beta}_{\mathrm{MDE}}$ in various flux-density windows $W(\varrho,J).$ Squares symbolize locations of window-centroids  $(\varrho+\Delta_\varrho/2,J+\Delta_J/2).$ The rectangle at the bottom right schematically visualizes the window-size used. Gray dots correspond to locations $(\rho_k,J_k)$ of individual data sub-samples before a separation procedure. Unstained areas indicate either a lack of data or belonging to a free-traffic regime. The plotted values are valid for a main lane. \label{fig:estimated-beta-MDE-main}}
\end{figure}

For completeness, we note that the results pictured in figures \ref{fig:LS-traffic}, \ref{fig:estimated-beta-MDE}, \ref{fig:estimated-alfa-MDE}, and \ref{fig:deviations-TCL-MDE} has been extracted from fast-lane data where car-car interactions are stronger (due to high speeds). Alternative results (valid for main-lane data) are visualized in figures \ref{fig:estimated-beta-MDE-main}, \ref{fig:estimated-alfa-MDE-main}, and \ref{fig:deviations-TCL-MDE-main} in the Appendix \ref{subsec:APP-03}. Comparing results obtained for fast and main lanes one can detect significant differences that are particularly noticeable in figures \ref{fig:estimated-beta-MDE} and \ref{fig:estimated-beta-MDE-main}. Here we detect different ranges of parameter $\beta$ and also different maxima for flux and density. All these disparities are caused predominantly by lower speeds of cars moving in a main lane.

Optimal parameters $\hat{\alpha},\hat{\beta}$ have been estimated by the method MLE as well. However, their values are (similarly to the results described in the subsection \ref{sec:DUE-LS}) extremely close to $\hat{\alpha}_{\mathrm{MDE}},\hat{\beta}_{\mathrm{MDE}}$ and therefore we will not depict them in figures.

\subsection{Damped matrices vs. freeway traffic -- direct link}

\begin{figure}[htbp]
\begin{center}
\hspace*{10mm}\includegraphics[width=9cm, angle=0]{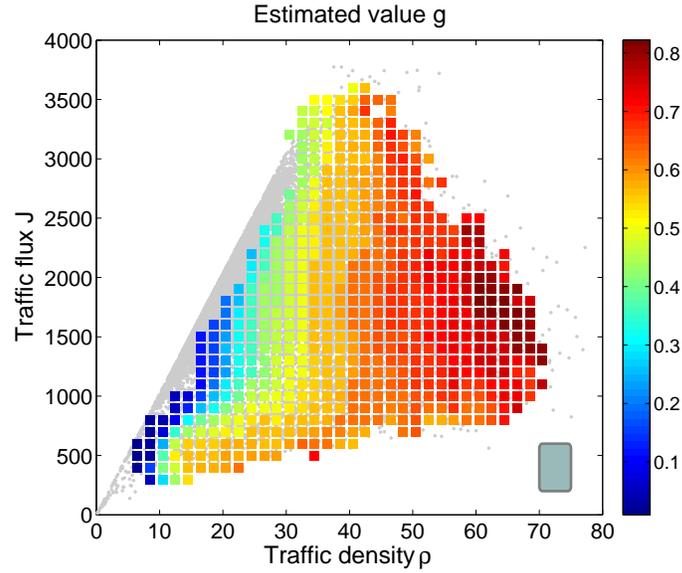}
\end{center}
\caption{Color representation of the estimated value $g$ obtained by the MDE-method. We plot values $\chi(\hat{\alpha}_{\mathrm{MDE}},\hat{\beta}_{\mathrm{MDE}})$ in various flux-density windows $W(\varrho,J).$ Squares symbolize locations of window-centroids  $(\varrho+\Delta_\varrho/2,J+\Delta_J/2).$ The plotted values are valid for a fast lane. \label{fig:gecko-optimalni}}
\end{figure}

\begin{figure}[htbp]
\begin{center}
\hspace*{10mm}\includegraphics[width=9cm, angle=0]{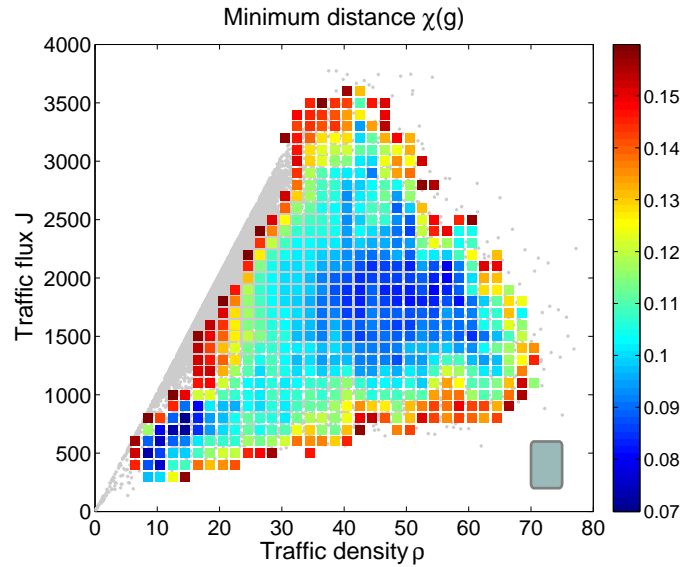}
\end{center}
\caption{Color representation of a minimum statistical distance $\chi(\alpha,\beta)$ obtained by the MDE-method. We plot values $\chi(\hat{\alpha}_{\mathrm{MDE}},\hat{\beta}_{\mathrm{MDE}})$ in various flux-density windows $W(\varrho,J).$ Squares symbolize locations of window-centroids  $(\varrho+\Delta_\varrho/2,J+\Delta_J/2).$ The plotted values are valid for a fast lane. \label{fig:gecko-minimalni-vzdalenosti}}
\end{figure}

Finally, we analyze a direct link between the random matrices investigated and traffic-flows monitored on expressways. For these intentions we minimize (analogously to the previous approaches) the statistical distance $\chi$ calculated for a histogram of vehicular clearances and for a histogram of level spacings in $\DUE-$ensembles. Thus, we search an optimal value of the damped parameter $g$ for which that distance is minimum. Comparison of the above-referred histograms is illustratively visualized in the figure \ref{fig:LS-traffic} (see histograms and crosses). Optimal values of $g$ are plotted in figure \ref{fig:gecko-optimalni}. Associated minimum distance are plotted in figure \ref{fig:gecko-minimalni-vzdalenosti}. As understandable, the values visualized in figure \ref{fig:gecko-minimalni-vzdalenosti} are higher that those enumerated as a distance between real-road distributions and the functional family (\ref{TC-distribution}). The main reason is that the former represents one-parameter fitting, whereas the latter represents an estimation that is two-parametric.

\section{Highlights, discussion and conclusion remarks}

This article has two scientific layers: mathematical (dealing with new types of random matrices) and physical (investigating properties of inner structure of vehicular streams). Firstly, this paper builds on the pioneering work \cite{Bogomolny-CM-matrices} dealing with new classes of random matrices. It brings a substantially deeper view into a spectral structure of damped matrices than the article \cite{Bogomolny-CM-matrices} itself. Specifically, using theoretical approaches we predict an analytical formula for level density of $\DUE-$matrices, describe particularities of the unfolding-procedure, and estimate specific values of parameters used in a predictive formula for the LS-distribution. Parameters in a two-parametric family (\ref{LS-distribution}) are estimated with help of the minimum distance estimator and the standard Kolmogorov estimator.

Secondly, we show that the same mathematical formula may be used as a theoretically-substantiated prognosis for a time-clearance distribution describing statistics of time-gaps among consecutive vehicles in real-road traffic. Indeed, both the  estimation-procedures used (minimum distance estimation and maximum likelihood estimation) prove that the function (\ref{TC-distribution}) represents a suitable candidate for a time-clearance distribution. In this work, for the first time, the time-clearance distribution is monitored in various flux-density sub-windows inside the entire flux-density plane. Thus, the output represents a map of estimated values $\alpha$ and $\beta$ depending on traffic density and flux. Since the parameter $\beta$ admits a thermodynamic interpretation (as is discussed in \cite{Traffic_NV,My_Multiheadways}) the changes of the socio-physical
coefficient $\beta=\beta(\varrho,J)$ reflect a level of mental pressure which the drivers are under during a given traffic constellation. As is evident in figure \ref{fig:estimated-beta-MDE} for fast lanes and \ref{fig:estimated-beta-MDE-main} for main lanes, a driver's mental strain depends predominantly on actual density and only marginally on actual flow. Among other things, we can clearly detect here a process of traffic saturation accompanied by corresponding changes in a traffic-microstructure. More specifically, with increasing density the coefficient $\beta$ is growing, which is reflected in more systematic configurations of vehicles  (as a consequence of the fact that the variance of a clearance distribution drops). Another benefit for physics of traffic is a relatively small statistical distance (if compared with previously published analyzes) between real-road histograms and a theoretical prediction. It confirms a quality of the approach used.

Finally, it has been demonstrated that the new type of random matrix ensembles ($g-$parameterized class $\DUE_g$) may be used to describe statistical distributions of inter-vehicle gaps. To be specific, similarly to the Ref. \cite{Cuernavaca,Cuernavaca2,Deift} (where it is proven that time-gap distribution monitored for Cuernavaca buses is the same as the level spacing distribution of $\GUE-$matrices) we now find that a similar correspondence exists between cars on expressways and the ensemble $\DUE.$ For a chosen traffic state (described by flux and density) there exists an optimal value of the parameter $g$ which the LS-distribution corresponds for to time-clearance distribution measured on a expressway. Figure (\ref{fig:gecko-optimalni}) shows that the force-parameter $g$ is increasing with traffic density, which supports an intuitive notion on changes of interaction forces in thickening flows.

To be sure that the above-mentioned similarities between traffic and Random Matrix Theory are more substantial (than presented in our article) one should show that the detected correspondence remains valid also if investigating second order characteristics (statistical rigidity or number variance). This will be covered in continuing researches. Preliminary numerical analysis, however, shows that a link between vehicular samples and eigenvalues of random matrices is definitely not accidental.
   
\subsection*{Acknowledgments}

The authors would like to thank The Road and Motorway Directorate of the Czech Republic (\v Reditelstv\' i silnic a d\'alnic \v CR) for providing traffic data analyzed in this paper.

Research presented in this work was supported by the grant 15-15049S provided by Czech Science Foundation (GA \v CR). Partial support was provided by Czech Technical University in Prague within the internal project SGS15/214/OHK4/3T/14.

\section{Appendix}

\subsection{Convergence of the distribution (\ref{pilgrim})} \label{subsec:APP-01}

Let $\D$ be the space of all smooth functions with a bounded support (according to \cite{Vladimirov}) and $\D'$ be the class of generalized functions (linear and continuous functionals over a space $\D.$) Then for an arbitrary $\varphi(x)\in\D:$
\begin{eqnarray}
\lim_{\varepsilon\rightarrow0_+} \left(q_{\theta,\varepsilon}(x);\varphi(x)\right)= \lim_{\varepsilon\rightarrow0_+} \int_\R \frac{\zeta(\theta)}{\varepsilon} \Theta(\varepsilon-|x|)\exp[-\frac{\theta^2\varepsilon^2}{\varepsilon^2-x^2}] \,\varphi(x)\,\dx=\nonumber\\ \zeta(\theta) \lim_{\varepsilon\rightarrow0_+} \int_\R \Theta(1-|y|)\exp[-\frac{\theta^2}{1-y^2}] \,\varphi(\varepsilon y)\,\dy=\nonumber \\ \zeta(\theta)\varphi(0) \int_{-1}^1\exp[-\frac{\theta^2}{1-y^2}]\,\dy=\varphi(0)=(\delta(x);\varphi(x)).
\end{eqnarray}
Therefore $\lim_{\varepsilon\rightarrow0_+} q_{\theta,\varepsilon}(x) \stackrel{\D'}{=} \delta(x).$ Analogously $\lim_{\varkappa\rightarrow0_+} \frac{1}{\sqrt{2\pi}\varkappa} \e^{-\frac{x^2}{2\varkappa^2}}\stackrel{\D'}{=} \delta(x),$ for
\begin{eqnarray}
\lim_{\varkappa\rightarrow0_+}  \left(\frac{1}{\sqrt{2\pi}\varkappa} \e^{-\frac{x^2}{2\varkappa^2}};\varphi(x)\right)= \lim_{\varkappa\rightarrow0_+} \int_\R \frac{1}{\sqrt{2\pi}\varkappa} \e^{-\frac{x^2}{2\varkappa^2}}\,\varphi(x)\,\dx=\nonumber\\ \lim_{\varkappa\rightarrow0_+} \int_\R \frac{1}{\sqrt{2\pi}} \e^{-y^2/2}\,\varphi(\varkappa y)\,\dy=\varphi(0)=(\delta(x);\varphi(x)).
\end{eqnarray}
For completeness, we add that $\Theta(x)$ stands for the Heaviside step function.

\subsection{Normalization and scaling in the distribution (\ref{LS-distribution})} \label{subsec:APP-02}

Denoting $J(x,\vk)=\int_0^\infty \e^{-\frac{x^2}{4t}}\e^{-\vk^2t}t^\alpha\,\dt$ the normalization and scaling conditions for the density (\ref{LS-distribution}) transform into the equation
\BE J(x,\vk)=-\frac{1}{2\vk}\frac{\p J}{\p \vk}. \EE
We add that the substitutions $D=\vk^2,$ $\beta=x^2/4
$ have been applied for convenience. Using the identity $J(x,\vk)=\frac{1}{2^\alpha}(\frac{x}{\vk})^{1+\alpha}\K_{\alpha+1}(x\vk)$ one finds
\BE (2\vk^2-1-\alpha)\K_{\alpha+1}(x\vk)=-x\vk\K_{\alpha+1}^\prime(x\vk).\EE
By means of the asymptotical expression
\BE \K_m(z)z^m \stackrel{z\rightarrow 0_+}{\approx}  (2m-2)!! \left(1+\frac{2z}{2m-1}\right)^{m-\frac{1}{2}}\e^{-z}, \quad (2m\geq1) \EE
we obtain the scaling equation for small $x:$
\BE 2x\vk^3+\vk^2(1+2\alpha-x^2)-2x\vk(\alpha+1)-(\alpha+1)(2\alpha+1)=0. \EE
Near the infinity this equation can be simplified to the form
\BE 4\vk^2-2x\vk-(2\alpha+3)=0, \label{scaling-equality}\EE
which leads to a detection of the linear trend in the scaling constant
\BE D(\beta) \approx \beta + \frac{3}{2} + \alpha  \quad (\beta \rightarrow \infty). \label{linear-asymp} \EE
On contrary, with help of the approximating expression
\BE \K_m(z) \stackrel{z\rightarrow \infty}{\approx}  \sqrt{\frac{\pi}{2}}\e^{-z}\left(\frac{1}{\sqrt{z}}+\frac{4m^2-1}{8z^{3/2}}\right) \EE
we obtain the scaling condition for large $x:$
\BE 16x\vk^3+2\vk^2(3+8\alpha+4\alpha^2-4x^2)-x\vk(15+16\alpha+4\alpha^2)-(4\alpha^3+12\alpha^2+11\alpha+3)=0. \EE
Surprisingly, this formula leads (after a simplification) to the scaling formula (\ref{scaling-equality}) as well, and therefore to the same linear asymptotics  (\ref{linear-asymp}).

Above that, since the variant $\beta=0$ in (\ref{LS-distribution}) represents the well-known gamma distribution it is easy to prove that $\lim_{\beta \rightarrow 0_+} D(\beta)=\alpha+1$ and therefore the final scaling formula (phenomenologically corrected) reads (\ref{D-scaling-constant}). Finally, for fixed values $\alpha,$ $\beta,$ and $D$ we have (\ref{A-scaling-constant}).

\subsection{Minimum distance estimations for main lanes} \label{subsec:APP-03}

In this part of Appendix we picture results of minimum distance estimations applied to traffic data extracted from main lanes of the Prague Ring.

\begin{figure}[htbp]
\begin{center}
\hspace*{10mm}\includegraphics[width=9cm, angle=0]{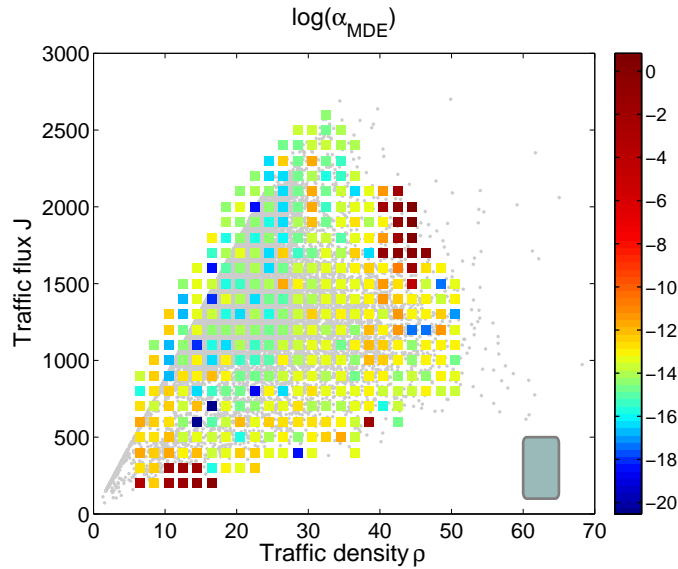}
\end{center}
\caption{Color representation of values $\alpha$ estimated by minimum distance estimation (\ref{MDE}). We plot values $\log(\hat{\alpha}_{\mathrm{MDE}})$ in various flux-density windows $W(\varrho,J).$ Squares symbolize locations of window-centroids  $(\varrho+\Delta_\varrho/2,J+\Delta_J/2).$ The plotted values are valid for a main lane. \label{fig:estimated-alfa-MDE-main}}
\end{figure}

\begin{figure}[htbp]
\begin{center}
\hspace*{10mm}\includegraphics[width=9cm, angle=0]{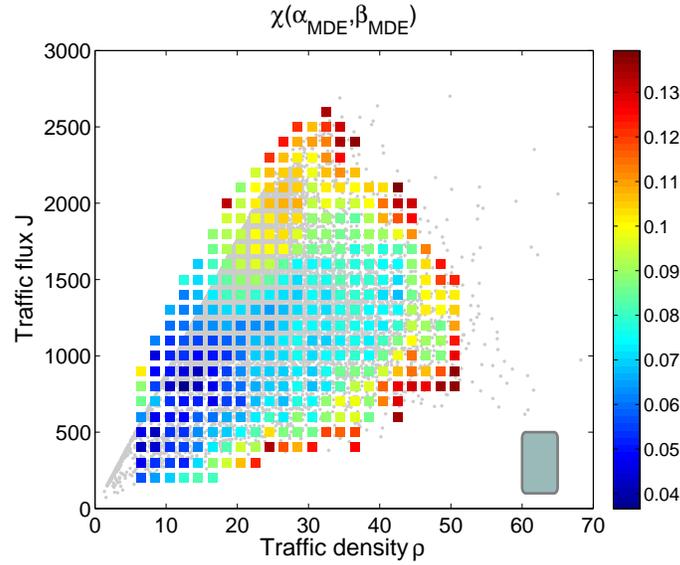}
\end{center}
\caption{Color representation of a minimum statistical distance $\chi(\alpha,\beta)$ obtained by the MDE-method. We plot values $\chi(\hat{\alpha}_{\mathrm{MDE}},\hat{\beta}_{\mathrm{MDE}})$ in various flux-density windows $W(\varrho,J).$ Squares symbolize locations of window-centroids  $(\varrho+\Delta_\varrho/2,J+\Delta_J/2).$ The plotted values are valid for a main lane. \label{fig:deviations-TCL-MDE-main}}
\end{figure}


\section*{References}

\begin{thebibliography}{10}

\bibitem{Tao}
Tao T 2014 \emph{Komplexit\"{a}t und Universalit\"{a}t} E-enterprise Lemgo

\bibitem{Cuernavaca}
Krb\'alek M and \v Seba P 2000 J. Phys. A: Math. Gen. \textbf{33} L229

\bibitem{Cuernavaca2}
Krb\'alek M and \v Seba P 2003 J. Phys. A: Math. Gen. \textbf{36} L7

\bibitem{Deift}
Baik J, Borodin A, Deift P and Suidan T 2006 J. Phys. A: Math. Gen. \textbf{39} (2006) 8965

\bibitem{Helbing_and_Krbalek}
Krb\'alek M and Helbing D 2004 Physica A \textbf{333} 370

\bibitem{Red_cars}
Krb\'alek M 2008 J. Phys. A: Math. Theor. \textbf{41} 205004

\bibitem{Traffic_NV}
Krb\'alek M and \v Seba P 2009 J. Phys. A: Math. Theor. \textbf{42} 345001

\bibitem{My_Multiheadways}
Krb\'alek M 2013 J. Phys. A: Math. Theor. \textbf{46} 445101

\bibitem{Red-intervals}
Krb\'alek M and \v Sleis J 2015 J. Phys. A: Math. Theor. \textbf{48} 015101

\bibitem{May}
May A D 1990 \emph{Traffic Flow Fundamentals} Englewood Cliffs, NJ: Prentice-Hall

\bibitem{Jin_and_Zhang} % Departure headways at signalized intersections: A log-normal distribution model approach
Jin X, Zhang Y, Wang F, Li L, Yao D, Su Y  and Wei Z 2009 Transportation Research Part C: Emerging Technologies \textbf{17(3)} 318

\bibitem{Li_and_Wang}%A new car-following model yielding log-normal type headways distributions
Li L, Wang F, Jiang R, Hu J and Ji Y 2010 Chinese Phys. B \textbf{19} 020513

\bibitem{Bogomolny-CM-matrices}
Bogomolny E, Giraud O and Schmit C 2011 Nonlinearity \textbf{24} 3179

\bibitem{Mehta}
Mehta M L 2004 \emph{Random matrices (Third Edition)} New York: Academic Press

\bibitem{Oxford}
Dodge Y 2003 \emph{The Oxford dictionary of statistical terms} New York: Oxford
University Press

\bibitem{Krbalek_gas}
Krb\'alek M 2007 J. Phys. A: Math. Theor. \textbf{40} 5813

\bibitem{K_means}
Hartigan A 1975 \emph{Clustering algorithms} John Wiley \& Sons

\bibitem{Martin}
Treiber M and Kesting A 2013 \emph{Traffic Flow Dynamics} Springer

\bibitem{Review-Kerner}
Kerner B S 2004 \emph{The Physics of Traffic} Berlin, New York: Springer Verlag


%%% APPENDIX

\bibitem{Vladimirov}
Vladimirov V. S. 1971 \emph{Equations of Mathematical Physics} New York: Dekker

\end{thebibliography}
\end{document}